\documentclass[]{aa}

\usepackage{graphicx}
\usepackage{natbib}
\usepackage{array}
\bibpunct{(}{)}{;}{a}{,}{,}

\begin{document}
\bibliographystyle{aa}

\title{The XMM-Newton RGS spectrum of the high luminosity\\
 Seyfert 1 galaxy Markarian 509}
\titlerunning{XMM-Newton RGS analysis of Markarian 509}
\author{R.A.N. Smith\thanks{E-mail: rans@mssl.ucl.ac.uk}, M.J. Page and G. Branduardi-Raymont}
\authorrunning{R.Smith, M.Page and G.Branduardi-Raymont}
\date{\today}
\institute{Mullard Space Science Laboratory, University College London, Holmbury St. Mary, Dorking, Surrey RH5 6NT, England}

\abstract{\\
We present a detailed analysis of the soft X-ray spectrum of the Seyfert 1 galaxy Markarian 509 taken with the \textit{XMM-Newton} Reflection Grating Spectrometer. An underlying power-law continuum and three warm absorber phases provide a good fit to the data along with a number of broad and narrow emission lines. Our three warm absorber phases each have different ionization parameters and column densities. We identify a low ionization, log\,$\xi$=0.89, high outflow velocity phase producing an Fe M-shell UTA feature along with absorption from O\,VI and N\,VI. There is an intermediate phase, log\,$\xi$=2.14, showing absorption from H-like carbon and nitrogen and He-like neon and oxygen. The third, high ionization, log\,$\xi$=3.26, low outflow velocity phase contains absorption from O\,VIII, Ne\,X and highly ionized iron. All phases are blueshifted with respect to the systemic velocity with flow velocities ranging from -60\,km\,s$^{-1}$ to -510\,km\,s$^{-1}$. The observed broad emission features have an RMS velocity of 8000$\pm$3000\,km\,s$^{-1}$ for the C\,VI and O\,VII lines.}

\maketitle

\section{Introduction}
\label{intro}

Soft X-ray absorption by ionized gas was first identified in Active Galactic Nuclei (AGN) by~\citet{halpern84} and recent observations indicate that over half of Seyfert 1 galaxies may have these `warm absorbers' (\nocite{reynolds97}Reynolds 1997, \nocite{george98}George et al. 1998). Since the launch of the \textit{XMM-Newton} and \textit{Chandra} high resolution grating spectrometers our ability to probe soft X-ray spectra of AGN has been greatly increased, enabling astronomers to do in-depth studies of the X-ray absorbing and emitting photoionized gas and to examine how they relate to the system as a whole (\nocite{blustin05}e.g. Blustin et al. 2005). Gas properties are examined to improve our understanding of the kinematics, ionization and composition of these systems. The warm absorber is often characterised by its ionization parameter, $\xi$. This determines the ionization state of the material and is defined as

\begin{equation}
\xi=\frac{L}{nr^2}
\end{equation}
where $L$ is the source luminosity (in erg\,s$^{-1}$), $n$ is the gas density (in cm$^{-3} $) and $r$ is the distance of the absorber from the centre of the system (in cm) (\nocite{tarter69}Tarter et al. 1969).

Markarian 509 was classified as a Seyfert 1 galaxy by~\citet{kopylov74} but is more like a Seyfert 1/quasi-stellar object (QSO) transition galaxy due to its compact appearance and high luminosity of $1.3-2.6\times 10^{44}\,$erg\,s$^{-1}$ in the 2-10\,keV range (\nocite{weaver01}Weaver et al. 2001). Its close proximity, z=0.034397 (\nocite{fisher95}Fisher et al. 1995), compared to other AGN of similar luminosity, and the low Galactic column density in the line of sight (N$_{H}=4.4\times10^{20}\,$cm$^{-2}$; \nocite{murphy96}Murphy et al. 1996), make it an ideal candidate for studying the inner regions of such objects.

Markarian 509 has well-studied intrinsic UV absorption (\nocite{kriss00}Kriss et al. 2000; ~\nocite{kraemer03}Kraemer et al. 2003) but much is still to be learnt from its high resolution X-ray spectrum. Markarian 509 was first identified as an X-ray source by~\citet{cooke78} and has since been observed with a number of instruments including \textit{EXOSAT} (\nocite{morini87}Morini et al. 1987), \textit{GINGA} (\nocite{singh90}Singh et al. 1990), and \textit{BeppoSAX} (\nocite{perola00}Perola et al. 2000).

\citet{pounds01} studied the \textit{XMM-Newton} Reflection Grating Spectrometer (RGS) spectrum of Markarian 509 taken in October 2000. They modelled it with a blackbody continuum as well as a weak Fe M-shell UTA region, Ne IX absorption and O\,VII narrow emission lines. ~\citet{yaqoob03} performed a thorough analysis of the \textit{Chandra} High-Energy Transmission Grating data. They  modelled the spectrum with two absorption edges from O\,VII and O\,VIII along with Ne\,IX, Ne\,X and O\,VIII line absorption. They fitted these spectral features using one phase of absorption through a photoionized gas with a column density 2.06$^{+0.39}_{-0.45}\times 10^{21}\,$cm$^{-2}$ and an ionization parameter of log\,$\xi=1.76^{+0.13}_{-0.14}$. The gas was found to be outflowing at $200\pm$170\,km\,s$^{-1}$ and a curve of growth analysis determined an approximate turbulent velocity of 100\,km\,s$^{-1}$.

~\citet{kriss00} analysed the UV spectrum of Markarian 509 obtained with the \textit{Far Ultraviolet Spectroscopic Explorer}. They modelled seven components for the UV absorber, which they grouped into two velocity phases with outflow velocities 0\,km\,s$^{-1}$ and 370\,km\,s$^{-1}$. They identified one component as having an ionization parameter and column density that could be compatible with an X-ray absorbing gas. With the exception of this component, however, they predicted that the UV and X-ray spectra are produced in separate zones within the AGN. ~\citet{kraemer03} analysed the UV spectrum of Markarian 509 using data from the Space Telescope Imaging Spectrograph echelle gratings on the \textit{Hubble Space Telescope}. They identified eight components in the UV absorber, with a similar range of velocities to Kriss et al. They did not find evidence for a component with an ionization and column density compatible with the X-ray absorbing phase. However, Kraemer et al. concluded from the kinematic association that the UV and X-ray absorbers may co-exist spatially, for example in the form of a two-phase clumpy medium.

In this paper we investigate the combined RGS spectral data from two observations by \textit{XMM-Newton}.  In Section~\ref{analysis} we describe the observations and data reduction, and the spectral fitting is reported in Section~\ref{fitting}. We discuss the results in Section~\ref{discussion} and present our conclusions in Section~\ref{conclusion}.

\section{Observations and Data Reduction}
\label{analysis}

Our data were obtained with the RGS (\nocite{denherder01}den Herder et al. 2001) onboard the \textit{XMM-Newton} observatory. This spectrometer provides unparalleled sensitivity in the 6\,--\,38\AA~(0.3\,--\,2.5\,ke\,V) range with an approximately constant resolution of 70\,m\AA~FWHM. The data analysed here are a combination of two observations of Markarian 509 by \textit{XMM-Newton}. The first took place on 25 October 2000 with an exposure of 23.4\,ks and was presented by~\citet{pounds01}. The second took place on 20 April 2001 with an exposure time of 30.6\,ks and the EPIC data were presented in \citet{page03} and \citet{dadina05}.

The data were processed using the \textit{XMM-Newton} Science Analysis System (SAS) version 5.2. First and second order spectra from both RGS, and both observations, were extracted. To correct for residual artefacts in the effective area calibration (\nocite{Stuhlinger05}Stuhlinger et al. 2005), the effective area of each response matrix was divided by the ratio of a powerlaw + Galactic column fit to the \textit{XMM-Newton} rev. 0084 RGS spectrum of the continuum source Markarian~421. The Markarian 509 spectra and response matrices were resampled and coadded to produce a single spectrum and a single response matrix. This method is explained in greater detail in~\citet{page03}, the only difference being that we do not treat the background separately as we use background subtracted RGS files. All of the nominal channel boundaries and energy ranges in the re-distribution matrix have been multiplied by 1+\,z in order to correct for redshift. To improve signal to noise the data were grouped by a factor of 3, resulting in a spectrum with $1000$ channels $\sim$\,30\,m\AA\ wide, well sampled with respect to the RGS resolution of $\sim$\,70\,m\AA\ FWHM. To correct for Galactic photoelectric absorption, the effective area elements of the response matrix were multiplied by the transmission as a function of energy for a column density of $4.1\,\times\,10^{20}\,$cm$^{-2}$. Finally, the wavelength scale was transformed to the rest frame of Markarian~509. In this way the data have been corrected for both astronomical redshift and absorption within our own Galaxy and so, from this point onwards, all data refer to the rest frame of Markarian 509. The main software used to analyse the data was \sc spex \rm 2.00, a spectral fitting package created by J. S. Kaastra with high resolution X-ray spectra in mind. 

\section{Spectral Fitting}
\label{fitting}
\subsection{Initial Analysis of the Data}

\begin{figure}
\includegraphics[scale=0.34, angle=270]{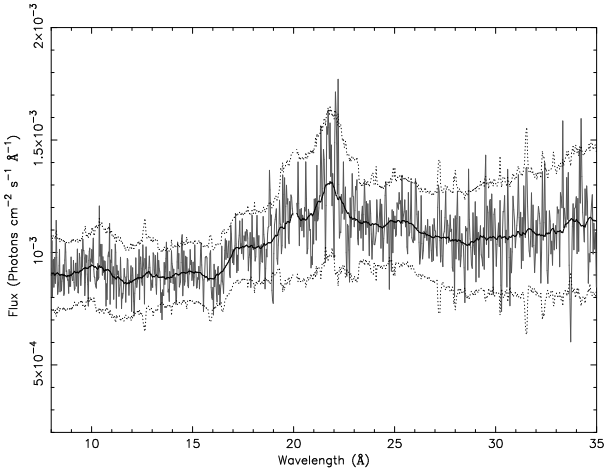}
\caption{The RGS spectrum of Markarian 509 (grey), the smoothed continuum used by the line finding program (black) and the 2$\sigma$ error bounds (dotted), in the rest frame of the galaxy.}
\label{smooth}
\end{figure}

\begin{figure}
\includegraphics[scale=0.34, angle=270]{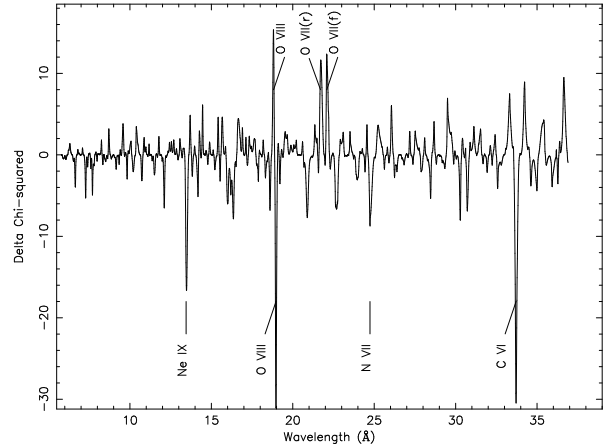}
\caption{A graph of $\Delta \chi^2$ against wavelength to show the significant lines in the spectrum, in the rest frame of the galaxy.}
\label{chi}
\end{figure}

Certain regions of the spectrum of Markarian 509 (Fig.~\ref{smooth}) are extremely complex, such as those between 12\AA~and 17\AA~and between 20\AA~and 24\AA. In these areas it can be difficult to distinguish absorption from emission features, with the risk that some features may be identified incorrectly or missed completely. Therefore, in order to get an estimate of the statistical significances of the major absorption or emission lines, the data were put through a program that identifies all of the significant spectral lines present (\nocite{page03a}Page et al. 2003a). Firstly a smooth continuum model is determined by averaging with a sliding cell, $\sim$\,1\AA~wide, and excluding all points more than 2$\sigma$~deviant from the smoothed continuum model (Fig.~\ref{smooth}). It is seen that the $2\sigma$ error bound has a number of sharp spikes in it, which are due to chip gaps in the instrument. Fortunately none of these occur at a point in the spectrum where there is an important spectral feature. A grid search is then performed adding a Gaussian of varying width, in steps of 50\,km\,s$^{-1}$ up to $\sigma=1000$\,km\,s$^{-1}$, to the continuum model at each point in the grid and calculating the goodness of fit between the model and the data by minimising $\chi^2$ (\nocite{lampton76}Lampton et al. 1976). This method is justified as the average number of counts in each bin is 140. This provides a measure of significance for all of the spectral features, shown in a graph of $\Delta\chi^2$ against wavelength in Fig~\ref{chi}.

The smoothed spectrum gives us a good overall view of the continuum shape; from this we are able to see the main features present, such as a number of possible broad emission lines, the most prominent being centred on $\sim$\,21.7\AA. We also see a rise in the continuum at $\sim$\,17\AA~possibly due to either the O\,VII edge or an Fe M-shell UTA. We now look to the $\Delta \chi^2$ graph where a number of strongly significant lines emerge. A $\Delta \chi^2$ value of 9 corresponds to a 3$\sigma$~significance for 1 interesting parameter. However, if one is scanning in wavelength space and the equivalent width of a feature is fitted, then one has to take into account the number of trials when assessing the significance of any features. Since we have approximately 450 independent resolution elements, we would expect 1.35 statistical fluctuations deviating at or more than $3\sigma$. There are three absorption lines with greater than a 3$\sigma$ significance at approximately 13.5\AA, 19.0\AA~and 33.7\AA. These correspond to the absorption lines of Ne\,IX, O\,VIII and C\,VI respectively. There are also some weaker features at 16.5\AA, from a possible Fe M-shell UTA, and at 24.7\AA, due to N\,VII. The Galactic O\,VII absorption line at  21.6\AA, seen here in the Markarian 509 rest frame at 20.9\AA, will be removed when fitting the spectrum at a later stage. Three emission features are also present at 18.8\AA, 21.7\AA~and 22.1\AA, which are likely to be due to O\,VIII and O\,VII emission. 

By taking a close look at the wavelengths of all of these lines we can calculate a flow velocity by measuring their shifts relative to the systemic velocity. These velocities, with the 2$\sigma$ error bounds, are shown in Table~\ref{chivel} along with the assumed species and the observed and emitted wavelengths.

\setlength{\extrarowheight}{0.1cm}
\begin{table}
\caption{Observed (in the rest frame of Markarian 509) and restframe wavelengths, and flow velocities derived from measuring the wavelengths of the individual lines.}
\begin{tabular}{|l|c|c|c|}
\hline
Species & $\lambda_{o}$ (\AA) & $\lambda_{e}$ (\AA) & Vel (km\,s$^{-1}$)\\
\hline
Ne\,IX absorption & 13.47$^{+0.03}_{-0.02}$ & 13.45 & $ +570^{+670}_{-450}$\\[3pt]
O\,VIII absorption & 18.97$^{+0.02}_{-0.02}$ & 18.97 & $ +10^{+160}_{-160}$\\[3pt]
N\,VII absorption & 24.74$^{+0.11}_{-0.04}$ & 24.78 & $ -440^{+850}_{-240}$\\[3pt]
C\,VI absorption & 33.71$^{+0.02}_{-0.02}$ & 33.73 & $ -210^{+180}_{-180}$\\[3pt]
O\,VIII emission & 18.82$^{+0.02}_{-0.03}$ & 18.97 & $ -2400^{+160}_{-320}$\\[3pt]
O\,VII(r) emission & 21.72$^{+0.07}_{-0.05}$ & 21.80 & $ -1100^{+690}_{-410}$\\[3pt]
O\,VII(f) emission & 22.09$^{+0.07}_{-0.03}$ & 22.10 & $ -170^{+540}_{-440}$\\[3pt]
\hline
\end{tabular}
\label{chivel}
\end{table}

\subsection{The Continuum and Broad Emission Lines}
\label{cont}

Now that we have the basic overall spectral profile and some approximate velocities we can go on to study the spectrum in more detail. Before looking at the absorption profile we first need to get a good fit to the underlying continuum. We start by fitting the data with a power law; however, there are regions where this is inadequate (for $\chi^2$ values see Table~\ref{chi_vary}). A number of broad bumps are apparent in the spectrum, as seen in Fig.~\ref{smooth}, and we find an improved match to the data by fitting a number of broad Gaussian emission lines. The spectrum in Fig.~\ref{smooth} shows broad peaks at 19.9\AA, 21.7\AA~and 25.2\AA. There also appears to be a narrow emission feature either side of the C\,VI absorption line at 33.7\AA~(Fig.~\ref{chi}), which can be fitted with another broad Gaussian line set to the wavelength of the C\,VI line.

\setlength{\extrarowheight}{0.1cm}
\begin{table*}
\begin{center}
\caption{The variation in $\chi^2$ for the different spectral models used; PL = Power law.}
\begin{tabular}{|l|c|c|c|c|}
\hline
Model & $\chi^2$ & Degrees of Freedom & Probability & PL Photon Index\\
\hline           
PL only & 1553 & 992 & $1.84\times10^{-27}$ & $2.19^{+0.01}_{-0.01}$\\[3pt]
PL + 2 broad Gaussians & 1463 & 988 & $4.72\times10^{-21}$ & $2.17^{+0.01}_{-0.01}$\\[3pt]
PL + 2 broad + 2 narrow Gaussians & 1416 & 984 & $3.93\times10^{-18}$ & $2.17^{+0.01}_{-0.01}$\\[3pt]
PL + 1 \sc xabs \rm components + 2 broad + 2 narrow Gaussians & 1116 & 980 & $1.52\times10^{-3}$ & $2.13^{+0.01}_{-0.02}$\\[3pt]
PL + 2 \sc xabs \rm components + 2 broad + 2 narrow Gaussians & 1074 & 976 & $1.52\times10^{-2}$ & $2.12^{+0.01}_{-0.01}$\\[3pt]
PL + 3 \sc xabs \rm components + 2 broad + 2 narrow Gaussians & 1018 & 972 & $1.51\times10^{-1}$ & $2.14^{+0.02}_{-0.01}$\\[3pt]

\hline
\end{tabular}
\label{chi_vary}
\end{center}
\end{table*}

\begin{figure*}
\begin{center}
\includegraphics[scale=0.7, angle=270]{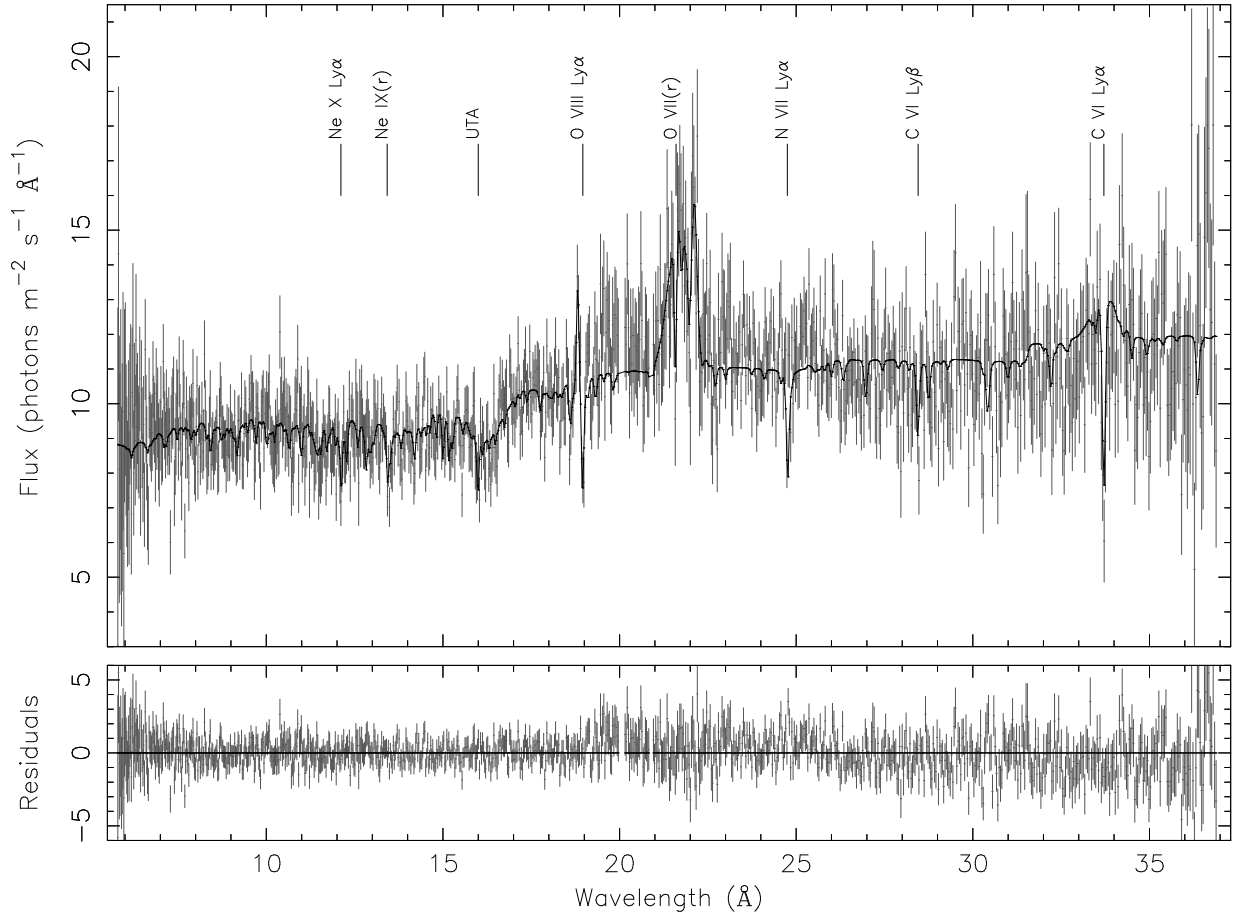}
\caption{The data fitted with three \sc xabs \rm components, two broad and two narrow Gaussians, and a power law, shown with the residuals (bottom graph).}
\label{res}
\end{center}
\end{figure*}

The broad emission lines at 21.7\AA, due to O\,VII, and 33.7\AA, due to C\,VI, give an average intrinsic RMS velocity of $8000\pm$ 3000\,km\,s$^{-1}$ (Table~\ref{emisparams}). We identify the 25.2\AA~broad feature as a possible C\,VI radiative recombination continuum (RRC; when electrons recombine with the ions, emitting photons). However, the identifications of this and the 19.9\AA~feature are not sufficiently reliable to be retained in the spectral model, which comprises a power law and two broad Gaussians at 21.7\AA~and 33.7\AA. The fitted continuum level at 20\AA~and 25\AA~will remain lower than the data, as can be seen from the residual graph in Fig.~\ref{res}; however, as we shall see in Section~\ref{abspn}, there are no significant absorption features in these regions, so this should not affect our analysis of the absorption profile significantly.

\setlength{\extrarowheight}{0.1cm}
\begin{table*}
\begin{center}
\caption{Emission line parameters for both the broad (top) and narrow (bottom) fitted lines. No flow velocity is given for the O\,VII broad emission line because it is a combination of the three lines of the triplet and so an accurate blueshift cannot be calculated. The wavelength and velocities are in the rest frame of the galaxy.}
\begin{tabular}{|c|c|c|c|c|c|}
\hline
Wavelength & FWHM & Norm & RMS Velocity & Flow Velocity & Transition\\
(\AA) & (\AA) & ($10^{50}$ph\,s$^{-1}$) & (km\,s$^{-1}$) & (km\,s$^{-1}$) &\\
\hline
$21.67^{+0.19}_{-0.18}$ & $0.63^{+0.32}_{-0.21}$ & $11.0^{+2.1}_{-2.0}$ & $8700^{+1800}_{-2000}$ & & O\,VII triplet\\[3pt]
$33.72^{+0.29}_{-0.31}$ & $0.80^{+0.55}_{-0.48}$ & $4.0^{+2.4}_{-2.3}$ & $7100^{+4000}_{-3900}$ & $-140^{+1800}_{-2400}$ & C\,VI Ly$\alpha$\\[3pt]
\hline
$18.82^{+0.10}_{-0.12}$ & $0^{+0.33}_{-0}$ & $1.1^{+0.5}_{-0.4}$ & $0^{+3200}_{-0}$ & $-2400^{+250}_{-680}$ & O\,VIII Ly$\alpha$\\[3pt]
$22.12^{+0.13}_{-0.12}$ & $0.13^{+0.26}_{-0.13}$ & $2.4^{+0.7}_{-0.6}$ & $1700^{+2500}_{-1900}$ & $+360^{+550}_{-310}$ & O\,VII(f)\\[3pt]
\hline
\end{tabular}
\label{emisparams}
\end{center}
\end{table*}

\subsection{Narrow Emission Lines}

Fig.~\ref{chi} indicates that there are a number of significant narrow emission lines present in the spectrum. The two emission features either side of the C\,VI absorption line have been identified as a broad C\,VI emission line, and the 21.7\AA, O\,VII(r), narrow emission line is confused by the presence of a broad O\,VII emission feature, so can no longer be reliably characterised. The results of a fit with two broad (O\,VII and C\,VI) and two narrow (O\,VIII and O\,VII(f)) emission lines are shown in Tables~\ref{chi_vary} and~\ref{emisparams}. 

The fitted emission lines have shifted wavelengths relative to the systemic velocity. These can be compared with those calculated from the line-finding program, without including any broad line features (Table~\ref{chivel}). If our identification of the emission lines is correct, the fit implies a flow velocity of approximately +360$\pm$400\,km\,s$^{-1}$ (Table~\ref{emisparams}) for the O\,VII(f) narrow emission line, at variance with the directly measured flow of -170$\pm$480\,km\,s$^{-1}$ (Table~\ref{chivel}). The O\,VIII narrow emission line is found to have a flow velocity of -2400$\pm$500\,km\,s$^{-1}$, in agreement with the directly measured results.

\begin{figure*}
\begin{center}
\includegraphics[keepaspectratio=false, angle =270, width=500pt, totalheight=500pt]{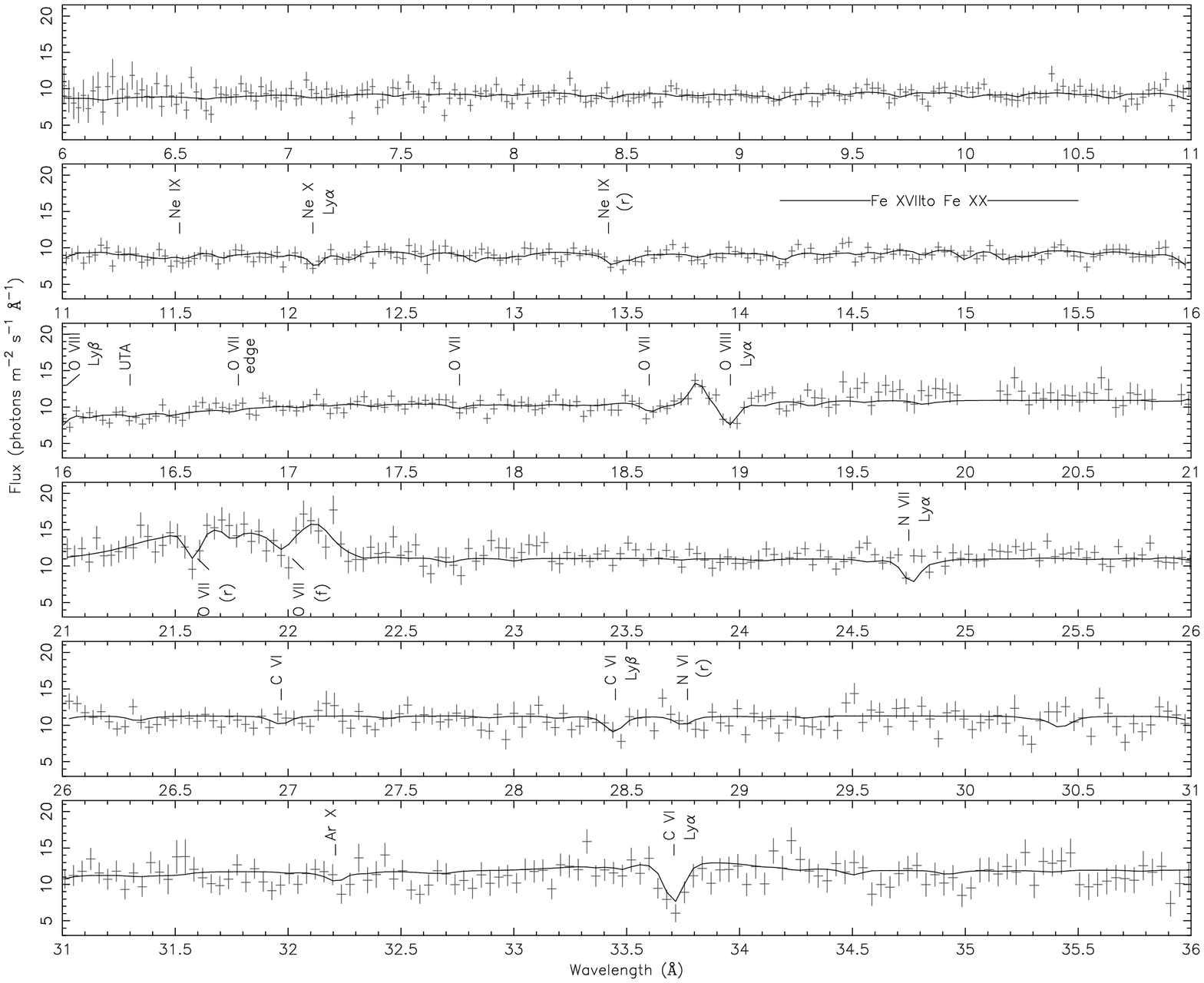}
\caption{A large scale view of Figure~\ref{res}: the fit includes a power law, three \sc xabs \rm components, two broad Gaussians for the broad O\,VII and C\,VI emission lines and two narrow Gaussians for the narrow O\,VIII and O\,VII(f) emission lines.}
\label{parts}
\end{center}
\end{figure*}

\subsection{Absorption Profile}
\label{abspn}

There are a number of spectral components available in \sc spex \rm that can now be added to the model to investigate the absorption lines that are present. The first model used here to analyse absorption is \sc slab\rm, which takes individual ionic column densities, an outflow velocity and a RMS velocity and computes the absorption profile for each ion.~\sc spex \rm varies the parameters to search for the best fit to the data. The outputs are then the best fit values for the column density, outflow velocity and RMS velocity for each ion. These velocities give an initial indication as to whether the entire warm absorber is kinematically associated or whether there are separate regions moving with different velocities.

In addition to the power law, two narrow and two broad Gaussian emission lines, we fitted individual absorption profiles for C\,VI, N\,VII, O\,VII, O\,VIII, Ne\,IX and Ne\,X (Table~\ref{slab40}), to our data. These ions were included because they produce the most significant lines in the spectrum (Fig.~\ref{chi}).~\citet{kriss00} find that one phase in their UV absorber has an ionization parameter log\,$\xi$=1.06, and column density 5$\times 10^{20}$cm$^{-2}$, comparable with an X-ray absorbing gas. Therefore, we used the velocity profile of this phase as a starting point for our analysis setting the RMS velocity to that found for this UV phase: 40\,km\,s$^{-1}$.

\setlength{\extrarowheight}{0.1cm}
\begin{table}
\caption{The parameters of the individually fitted \sc slab \rm absorption components, using a fixed RMS velocity value of 40\,km\,s$^{-1}$, with 1$\sigma$ error bounds. The flow velocity is in the rest frame of the galaxy.}
\begin{tabular}{|l|c|c|}
\hline
Species & Column Density & Flow Velocity\\
 & log\,N (cm$^{-2}$) & (km\,s$^{-1}$)\\
\hline
C\,VI &$ 17.38^{+0.15}_{-0.16} $&$ -230^{+150}_{-130}$\\[3pt]
N\,VII &$ 17.98^{+0.09}_{-0.06} $&$ -240^{+200}_{-180}$\\[3pt]
O\,VII &$ 17.80^{+0.07}_{-0.10} $&$ -210^{+110}_{-250}$\\[3pt]
O\,VIII &$ 18.13^{+0.07}_{-0.6} $&$ -10^{+120}_{-110}$\\[3pt]
Ne\,IX &$ 17.40^{+0.16}_{-0.18} $&$ 0^{+650}_{-360}$\\[3pt]
Ne\,X &$ 17.63^{+0.23}_{-0.27} $&$ 60^{+110}_{-820}$\\[3pt]
\hline
\end{tabular}
\label{slab40}
\end{table}

\begin{figure}
\includegraphics[angle=270,scale=0.37]{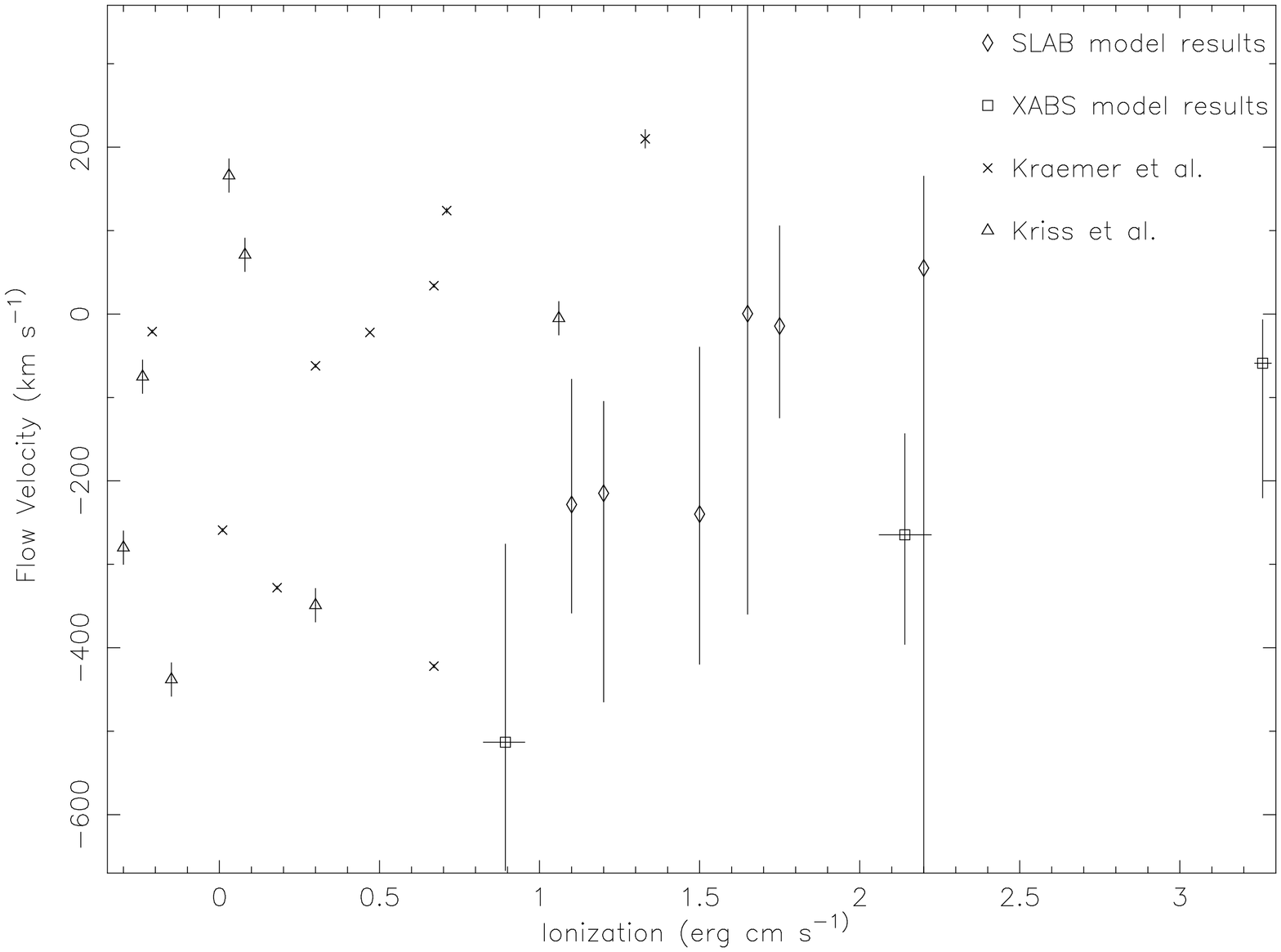}
\caption{The relationship between the ionization parameter at maximum abundance and flow velocity of the elements in our \sc slab \rm and \sc xabs \rm models. Shown also are the UV results from~\citet{kriss00} (with conservative uncertainties estimated from their spectral resolution) and~\citet{kraemer03} including their given errors.}
\label{slabgraph}
\end{figure}

Some of the flow velocities are not well constrained making a detailed analysis difficult. However, we have plotted the velocities against the ionization parameter at maximum abundance to see if there is any connection between the two and to see how they relate to the UV results (Fig.~\ref{slabgraph}). From the \sc slab \rm results, in Table~\ref{slab40}, we see that there appear to be two separate regions with different flow velocities, one at approximately -230\,km\,s$^{-1}$ and one at the systemic velocity.

An alternative \sc spex \rm model we use is \sc xabs\rm. This provides a more physical view of the gas by providing a self-consistent model that contains all of the ions, in contrast to fitting each ion separately with \sc slab\rm.~\sc xabs \rm applies absorption by photoionized gas, with variable column density, ionization parameter, elemental abundances, RMS velocity and flow velocity, to the model. The ionization parameter determines which transitions are likely to occur in a given phase. By using this model we are able to see if there are multiple phases in the gas and ascertain the parameters for each phase.  

\begin{figure}
\includegraphics[scale=0.34, angle=270]{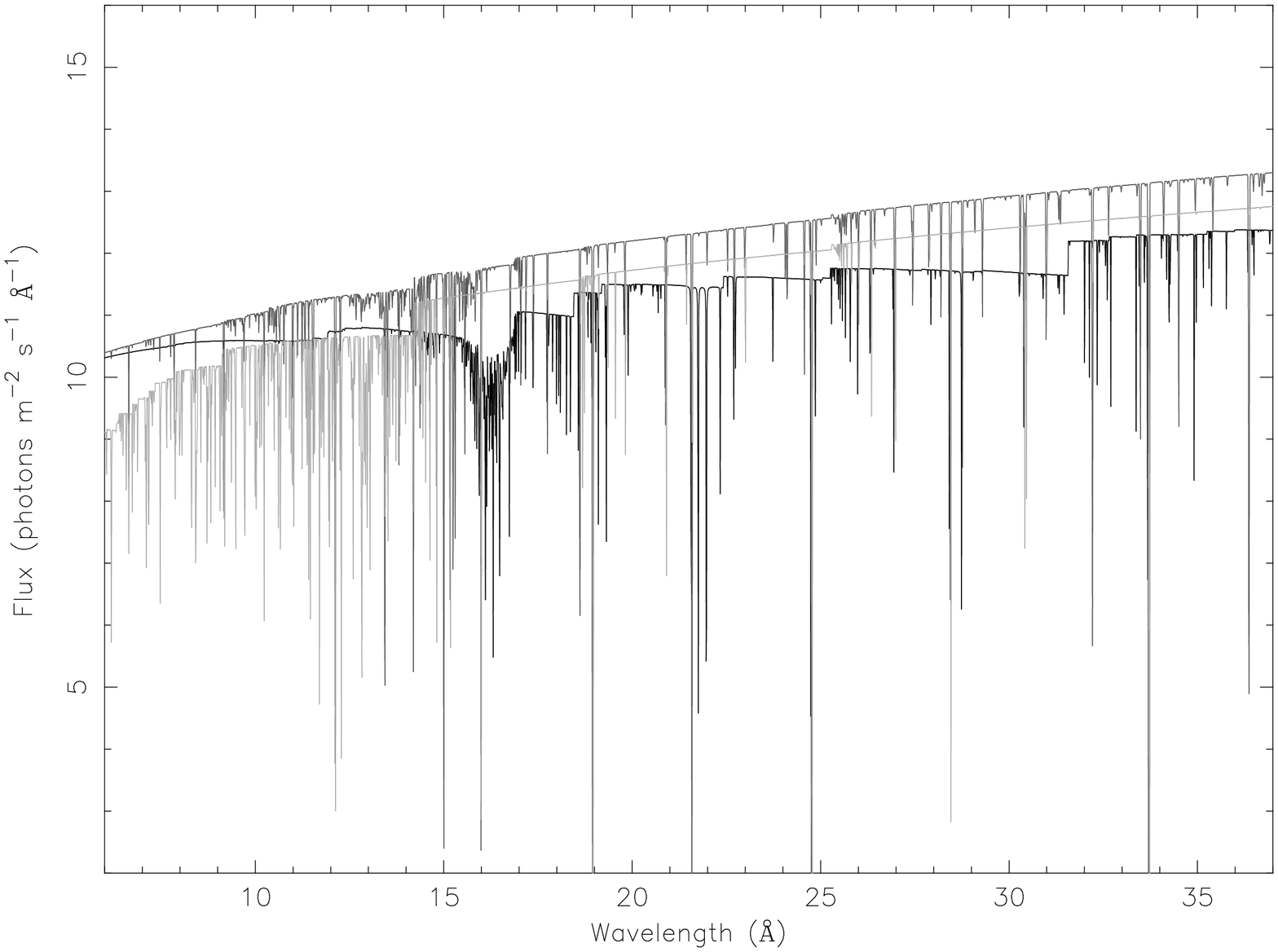}
\caption{A model plot showing the contribution to the spectrum by each of the three phases, the parameters of which are shown in Table~\ref{fixabs}. Phase 1 is shown in black, Phase 2 in dark grey and Phase 3 in light grey.}
\label{model}
\end{figure}

The data were fitted with different numbers of \sc xabs \rm models, the $\chi^2$ values for which are shown in Table~\ref{chi_vary}. The abundances of all elements were set to the solar values. We see from the table the improvement in the quality of the fit found by adding multiple \sc xabs \rm components to the model. We conclude that three \sc xabs \rm components provide a good match to the data. The best fit parameters (Table~\ref{fixabs}) suggest three distinct regions, each with a different ionization parameter and outflow velocity. By plotting each phase of the model separately (Fig.~\ref{model}) we can say which phase makes the most significant contribution to each absorption line. From the best fit parameters we see that there is a cool, high velocity phase producing the low ionization, log\,$\xi$=0.89, iron transitions forming the UTA at 16\AA, along with the N\,VI and O\,VI absorption lines. The intermediate phase, log\,$\xi$=2.14, produces the C\,VI, N\,VII, Ne\,IX and O\,VII absorption lines, whilst the low velocity, high ionization, log\,$\xi$=3.26, phase produces the Ne\,X, O\,VIII and highly ionized iron absorption features. 

\setlength{\extrarowheight}{0.1cm}
\begin{table*}
\begin{center}
\caption{The \sc xabs \rm best fit parameters for a model including a power law, two broad and two narrow Gaussian emission lines and three \sc xabs \rm components, with the 2$\sigma$ error bounds. The velocities are in the rest frame of the galaxy.}
\begin {tabular}{|l|c|c|c|c|l|}
\hline
Phase & Flow Velocity & RMS Velocity & log\,$\xi$ & Column Density & Prominent Ions\\
 & (km\,s$^{-1}$) & (km\,s$^{-1}$) & (erg\,cm\,s$^{-1}$) & ($10^{21}$cm$^{-2}$) & \\
\hline
Phase 1 & -510$^{+290}_{-140}$ & 0$^{+20}_{-0}$ & 0.89$^{+0.13}_{-0.11}$ & 0.79$^{+0.15}_{-0.13}$ & O\,VI, N\,VI, Fe\,IX--Fe\,XIII\\[3pt]
Phase 2 & -260$^{+120}_{-130}$ & 70$^{+70}_{-30}$ & 2.14$^{+0.19}_{-0.12}$ & 0.75$^{+0.19}_{-0.11}$ & O\,VII, N\,VII, C\,VI, Ne\,IX\\[3pt]
Phase 3 & -60$^{+70}_{-200}$ & 0$^{+30}_{-0}$ & 3.26$^{+0.18}_{-0.27}$ & 5.5$^{+1.3}_{-1.4}$ & O\,VIII, Ne\,X, Fe\,XVII--Fe\,XX\\[3pt]
\hline
\end{tabular}
\label{fixabs}
\end{center}
\end{table*}

Figs.~\ref{res} and~\ref{parts} show the best fit, which includes a power law, two broad emission lines, at 21.8\AA~and 33.7\AA, two narrow emission lines, at 18.9\AA~and 22.1\AA, and three \sc xabs \rm components with the parameters listed in Table~\ref{fixabs}. In addition to the elements mentioned above there may also be contributions to the absorption profile from Silicon, Magnesium, and Argon. However, their impact on the spectrum is so small that they cannot be discussed in detail in this analysis. 

\subsection{Ionization Structure}

Using the best fit values obtained in Section~\ref{abspn} we are able to determine whether the three phases of the absorber are in pressure equilibrium (\nocite{krolik81}Krolik et al. 1981). In order to do this we use the pressure form of the ionization parameter ($\Xi$), which is related to $\xi$ by

\begin{equation}
\Xi=\frac{L}{4\pi cr^2 P}=\frac{0.961\times 10^4 \xi}{T}
\end{equation}
where $L$ is the luminosity, $r$ is the distance of the absorber from the ionizing source, $P$ is the pressure and $T$ the temperature (\nocite{steenbrugge05}Steenbrugge et al. 2005).

\begin{figure}
\includegraphics[scale=0.34, angle=270]{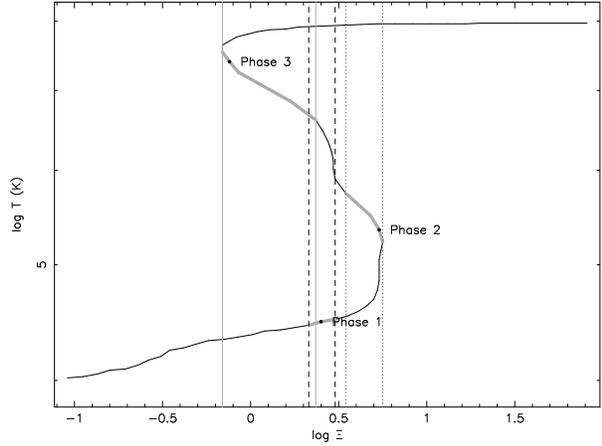}
\caption{Thermal stability showing the equilibrium temperature as a function of $\Xi$ for the SED described in~\cite{yaqoob03}. The positions relating to the three warm absorber phases are indicated along with their errors, displayed as the thick lines. For ease of comparison the errors have also been indicated with vertical lines; Phase 1 shown with black dashed lines, Phase 2 with dark grey dotted lines and Phase 3 with light grey solid lines.}
\label{equil}
\end{figure}

There is a specific range in ionization for which pressure equilibrium is possible (\nocite{krolik01}Krolik et al. 2001). In order to investigate this we used the \sc xstar \rm ionization code and the SED shown in~\cite{yaqoob03} to produce Fig.~\ref{equil}. Regions where $dT/d\Xi>0$ are stable against isobaric thermal perturbations, with regions where $dT/d\Xi<0$, i.e. where the curve doubles back, being unstable. In order for the different phases to be in pressure equilibrium they must have the same value of $\Xi$. Included in Fig.\ref{equil} are the locations of the three phases of the warm absorber in Markarian 509. From this it is possible to see that the errors in $\Xi$ on Phase 1 and 3 marginally overlap. However, if Phase 3 had a compatible $\Xi$ value with Phase 1 then it would be on the unstable branch and so would be unlikely to remain there. Physically Phases 1 and 2 are more likely to be in pressure equilibrium because they are more stable and have similar values of $\Xi$. Therefore, it is unlikely that all three phases identified here coexist as part of the same structure, but Phases 1 and 2 could be in pressure equilibrium.

\section{Discussion}
\label{discussion}
\subsection{Broad Emission Lines}

We have identified, and included in the fits, two broad emission lines due to O\,VII at 21.7\AA, and C\,VI at 33.7\AA, in the RGS spectrum of Markarian 509. By fitting the lines with \sc spex \rm an RMS velocity value of 8700$\pm2000$\,km\,s$^{-1}$ was found for the O\,VII line and 7100$\pm4000$\,km\,s$^{-1}$ for the C\,VI line. We also identify a possible C\,VI RRC at 25.2\AA, similar to that seen in NGC4051 (\nocite{ogle04}Ogle et al. 2004), compatible with the C\,VI broad line. There is also a broad peak at 19.9\AA: possible explanations for this include a N\,VII absorption edge (18.6\AA) or an Fe\,XXIII broad emission line (19.9\AA). However, for the N\,VII absorption edge to be large enough to account for the rise in the continuum the two fitted N\,VII absorption lines would have to be approximately 50\% deeper than is observed. Fe\,XXIII is also unlikely since we would expect to see other, stronger lines emitted from this species at 8.3\AA~and 10\AA. The 25.2\AA~and 19.9\AA~features have not been included in the fits because they are not firmly identified with any strong expected transitions.

The RMS velocity values of the C\,VI broad emission line suggests that it does not originate in any phase of the warm absorber, but is likely to arise in the broad line region of the AGN close to the centre of the system. We have not given a flow velocity value for the O\,VII broad emission line as it is likely to be a combination of the three lines of the O\,VII triplet making any velocity measurement unreliable. This will also alter the fitted RMS velocity value; we only fit one broad Gaussian to the triplet. Therefore, each of the three emission lines is likely to have a smaller RMS value than that shown, but must still be broad since we only see one line.

Broad emission lines, similar to those found here, have been seen in the soft X-ray spectra of other Seyfert 1 galaxies such as NGC 5548 (\nocite{steenbrugge05}Steenbrugge et al. 2005), Markarian 279 (\nocite{costantini05}Costantini et al. 2005) and 3C273 (Page et al. in prep.). Steenbrugge et al. identify both the O\,VII, 21.7\AA, and C\,VI, 33.7\AA, broad emission lines, which are also our two most reliable broad lines. The lines in NGC 5548 are fitted with a FWHM velocity value of 8000\,km\,s$^{-1}$, taken from the corresponding UV broad lines. ~\citet{kriss00} report broad O\,VI and C\,III emission lines with a FWHM of 11000\,km\,s$^{-1}$ in the UV spectrum of Markarian 509, a good match with our broad X-ray emission lines. 

\subsection{Narrow Emission Lines}
 
If we identify the 18.8\AA~line as an O\,VIII emission line, then it has an outflow velocity of 2400$\pm$500\,km\,s$^{-1}$. This is a much larger flow velocity than any other narrow emission line in the X-ray or UV spectra of Markarian 509. However, the O\,VIII absorption line, at 18.9\AA, is likely to remove the long wavelength side of the emission line making it narrower and moving the peak to an apparently shorter wavelength. We find this a more probable answer than identifying the line as an inflowing, high order, O\,VII transition with a flow velocity of +3000\,km\,s$^{-1}$. 

The second narrow emission line included in our fit is the forbidden O\,VII line at 22.1\AA. This is part of the O\,VII triplet and so other lines are to be expected, but these have been overshadowed by the broad O\,VII emission line centred at 21.8\AA. Note the apparent flow velocities for our two narrow emission lines are significantly different (Table~\ref{emisparams}); this is likely to be due to both lines being confused by the presence of other emission and absorption lines and so it is still possible that these lines originate in the same region of the AGN and have the same velocity structure. This broad line contamination is also the likely cause of the discrepancy between the fitted (Table~\ref{emisparams}) and the directly measured (Table~\ref{chivel}) flow velocities for the O\,VII(f) narrow emission line.

\subsection{Characteristics of the Absorber}

In the spectrum of Markarian 509 we find three separate absorption phases: Phase 1 contributes the Fe M-shell UTA and the O\,VI and N\,VI absorption features, Phase 2 contains absorption from O\,VII, N\,VII, C\,VI and Ne\,IX, and Phase 3 produces the O\,VIII, Ne\,X and highly ionized iron absorption features (Table~\ref{fixabs}). We also identify minor contributions to the spectrum from Silicon, Magnesium and Argon. The confidence intervals for our ionization parameters have no overlap in log\,$\xi$, so we are looking at at least three distinct phases in the gas; however, a continuous ionization distribution throughout the gas cannot be excluded. We have shown that it would be possible for Phases 1 and 2 to coexist in the same structure as they have compatible velocity profiles and may have the same value of the pressure form of the ionization parameter, so could be in pressure equilibrium.

We have improved the reliability of our results by using a number of different techniques to analyse the absorption profile. Fig.~\ref{slabgraph} shows our \sc slab \rm and \sc xabs \rm results along with UV information from previous analyses. We can see that all of the individually fitted \sc slab \rm absorption lines have the same flow velocity, within the errors, as their main contributing phase, found from the \sc xabs \rm analysis. These results are also backed up by the flow velocities derived from the line-finding program (Table~\ref{chivel}). Some of the values shown are not well constrained due to the complexity of the spectrum and the frequent confusion between spectral features. This is especially evident in the case of the Ne\,IX and Ne\,X absorption lines, both of which are confused by the presence of highly ionized iron features, thereby decreasing the reliability of the fitted values. There is also the additional problem of absorption lines being produced in more than one phase (Fig.~\ref{model}) causing the apparent line width to increase and the apparent wavelength of the line to change. 

~\citet{kriss00} and~\citet{kraemer03} identify carbon and oxygen as the main contributors to the UV absorption spectrum and also find some nitrogen and silicon, all of which we have identified in the RGS X-ray spectrum. Interestingly, in Fig.~\ref{slabgraph} we see that a number of the UV absorbers identified have a similar outflow velocity to our Phase 1 and are also within the ionization range that this phase covers. We are able to predict the UV spectrum that our phases would produce and we find that Phase 1 contributes significantly to the carbon, oxygen and nitrogen features in the UV spectrum. We have used the ratios of the fluxes of these lines and the velocity profiles to find which phases are compatible and conclude that our Phase 1 is compatible with the UV Phase 2 from the Kraemer et al. and Kriss et al. analyses. Our Phase 2 also contributes to the UV spectrum but we have been unable to reach a definite conclusion about which UV phase it corresponds to.  

Kriss et al. identify one of their regions, Phase 5, as being compatible with an X-ray producing gas, having a column density of 5$\times 10^{20}$cm$^{-2}$, ionization of log\,$\xi$=1.06, an outflow velocity of 5\,km\,s$^{-1}$ and a FWHM of 40\,km\,s$^{-1}$. This ionization and column density are compatible with our X-ray Phase 1, but the velocity profiles do not correspond so we conclude that this particular UV phase is not compatible with any of our X-ray phases.

We can also look back to the previous analyses of this warm absorber and compare our findings. \citet{pounds01} observed a weak Fe M-shell UTA region, a Ne\,IX absorption line and the O\,VII narrow emission line triplet in the first~\textit{XMM-Newton} RGS spectrum of Markarian 509 taken on 25 November 2000. These features are observed in our spectrum apart from an identification of all three lines of the O\,VII triplet. The analysis of Pounds et al. finds no evidence of carbon or nitrogen absorption features and also appears to indicate an unusual ratio between the triplet lines (see~\citet{mewe03} for triplet line ratios). As we have shown, two of the triplet emission lines are not detected after the inclusion of the broad O\,VII emission line. Pounds et al. identify two separate phases producing the RGS spectrum but do not give any values for the ionization parameters or column densities for us to compare with our analysis.

\citet{yaqoob03} analysed Markarian 509 \textit{Chandra} data and fitted two absorption edges from O\,VII and O\,VIII. These features are not observed clearly in our data, which may be due to our better definition of the UTA region, mimicking the O\,VII absorption edge. The O\,VIII edge at 14.2\AA~is extremely weak and so is not a prominent feature in our RGS data. Yaqoob et al. also observed no clear evidence of an Fe M-shell UTA centred around 16\AA. However, there is good agreement in the identification of Ne\,IX and Ne\,X lines, as well as O\,VIII absorption, which are some of the strongest lines in our RGS spectrum as well. These lines were fitted by Yaqoob et al. with an ionization parameter of log\,$\xi=1.76^{+0.13}_{-0.14}$ and column density 2.06$^{+0.39}_{-0.45}\times 10^{21}$cm$^{-2}$, which lie between our Phase 1 and Phase 2 in ionization; however, they fitted only one phase of absorbing gas to the data and so the results are expected to differ somewhat. There is a good velocity agreement with our Phase 2, as Yaqoob et al. fit an outflow velocity of 200$\pm$170\,km\,s$^{-1}$ and use a curve of growth analysis to derive a velocity width of 100\,km\,s$^{-1}$. There will be differences between Yaqoob et al.'s and our results due to their data being taken with a higher spectral resolution and lower signal to noise ratio. Moreover, Markarian 509 has a variable luminosity (\nocite{jackson00}Jackson et al. 2000,~\nocite{carone96}Carone et al. 1996) with L$_{2-10\,keV}$ typically in the range (1.3-2.6)$\times 10^{44}$erg\,s$^{-1}$ (Weaver et al. 2001), which may alter the spectral profile over time. The 0.5-10\,keV unabsorbed flux during the first observation with \textit{XMM-Newton} was 2.6$\times 10^{-11}$erg\,cm$^{-2}$\,s$^{-1}$, a factor of 2.8 less than the 0.5-10\,k\,eV flux of 7.3$\times 10^{-11}$erg\,cm$^{-2}$\,s$^{-1}$ during the \textit{Chandra} observation. The second \textit{XMM-Newton} observation occurred just 6 days after the \textit{Chandra} observation with a 0.5-10\,k\,eV flux of 6.3$\times 10^{-11}$erg\,cm$^{-2}$\,s$^{-1}$.

\section{Conclusion}
\label{conclusion}

We have used the high resolution X-ray spectra obtained with the Reflection Grating Spectrometer onboard \textit{XMM-Newton} to study the characteristics of the Seyfert 1 galaxy Markarian 509 and present the most detailed analysis to date of the X-ray absorbing and emitting photoionized gas present in this system. 

The continuum has been fitted with a simple power law. The analysis suggests the presence of a warm absorber consisting of three phases: Phase 1 contains the high outflow velocity, low ionization O\,VI and N\,VI absorption lines along with an Fe M-shell UTA; Phase 2 contributes the medium ionization O\,VII, N\,VII, C\,VI and Ne\,IX absorption features, and Phase 3 produces the low outflow velocity, high ionization O\,VIII, Ne\,X and highly ionized iron absorption features. However, we cannot rule out the possibility of a continuous ionization change throughout one gas system. It is possible that Phases 1 and 2 may coexist in pressure equilibrium, but it is unlikely that Phase 3 is also part of the same structure. We have identified Phase 1 as being compatible with a UV phase identified in previous studies.

In addition to the absorption features identified here, there is also evidence of emission in the spectrum of Markarian 509. We identify two narrow emission lines due to O\,VII at 22.1\AA, and O\,VIII at 18.8\AA. We also identify two broad emission lines, due to O\,VII at 21.7\AA, and C\,VI at 33.7\AA, which have an RMS velocity of 8000$\pm$3000\,km\,s$^{-1}$ consistent with originating in the broad line region. 

\acknowledgements

The work shown here is based on observations obtained with \textit{XMM-Newton}, an ESA science mission with instruments and contributions directly funded by ESA Member states and the USA(NASA). R.A.S. acknowledges the support of a PPARC studentship.

\bibliography{rans}

\begin{thebibliography}{31}
\expandafter\ifx\csname natexlab\endcsname\relax\def\natexlab#1{#1}\fi
\expandafter\ifx\csname url\endcsname\relax
  \def\url#1{{\tt #1}}\fi
\expandafter\ifx\csname urlprefix\endcsname\relax\def\urlprefix{URL }\fi

\bibitem[{Blustin et~al.(2005)Blustin, Page, Fuerst, Branduardi-Raymont, \&
  Ashton}]{blustin05}
Blustin A.J., Page M.J., Fuerst S.V., Branduardi-Raymont G., Ashton C.E., 2005,
  A\&A, 431, 111

\bibitem[{{Carone} et~al.(1996){Carone}, {Peterson}, {Bechtold}
  et~al.}]{carone96}
{Carone} T.E., {Peterson} B.M., {Bechtold} J., et~al., Nov. 1996, ApJ, 471, 737

\bibitem[{{Cooke} et~al.(1978){Cooke}, {Ricketts}, {Maccacaro}
  et~al.}]{cooke78}
{Cooke} B.A., {Ricketts} M.J., {Maccacaro} T., et~al., Feb. 1978, MNRAS, 182,
  489

\bibitem[{{Costantini} et~al.(2005){Costantini}, {Kaastra}, {Steenbrugge}
  et~al.}]{costantini05}
{Costantini} E., {Kaastra} J.S., {Steenbrugge} K.C., et~al., Jun. 2005, In:
  {Smith} R. (ed.) AIP Conf. Proc. 774: X-ray Diagnostics of Astrophysical
  Plasmas: Theory, Experiment, and Observation, 321--323

\bibitem[{{Dadina} et~al.(2005){Dadina}, {Cappi}, {Malaguti}, {Ponti}, \& {de
  Rosa}}]{dadina05}
{Dadina} M., {Cappi} M., {Malaguti} G., {Ponti} G., {de Rosa} A., Nov. 2005,
  A\&A, 442, 461

\bibitem[{{den Herder} et~al.(2001){den Herder}, {Brinkman}, {Kahn}
  et~al.}]{denherder01}
{den Herder} J.W., {Brinkman} A.C., {Kahn} S.M., et~al., Jan. 2001, A\&A, 365,
  L7

\bibitem[{{Fisher} et~al.(1995){Fisher}, {Huchra}, {Strauss} et~al.}]{fisher95}
{Fisher} K.B., {Huchra} J.P., {Strauss} M.A., et~al., Sep. 1995, ApJ, 100, 69

\bibitem[{{George} et~al.(1998){George}, {Turner}, {Netzer} et~al.}]{george98}
{George} I.M., {Turner} T.J., {Netzer} H., et~al., Jan. 1998, ApJS, 114, 73

\bibitem[{{Halpern}(1984)}]{halpern84}
{Halpern} J.P., Jun. 1984, ApJ, 281, 90

\bibitem[{{Jackson} et~al.(2000){Jackson}, {Leighly}, \&
  {Matsuoka}}]{jackson00}
{Jackson} M.S., {Leighly} K.M., {Matsuoka} M., Dec. 2000, Bulletin of the
  American Astronomical Society, 32, 1457

\bibitem[{{Kopylov} et~al.(1974){Kopylov}, {Lipovetskii}, {Pronik}, \&
  {Chuvaev}}]{kopylov74}
{Kopylov} I.M., {Lipovetskii} V.A., {Pronik} V.I., {Chuvaev} K.K., Nov. 1974,
  Astrofizika, 10, 483

\bibitem[{{Kraemer} et~al.(2003){Kraemer}, {Crenshaw}, {Yaqoob}
  et~al.}]{kraemer03}
{Kraemer} S.B., {Crenshaw} D.M., {Yaqoob} T., et~al., Jan. 2003, ApJ, 582, 125

\bibitem[{{Kriss} et~al.(2000){Kriss}, {Green}, {Brotherton} et~al.}]{kriss00}
{Kriss} G.A., {Green} R.F., {Brotherton} M., et~al., Jul. 2000, ApJ, 538, L17

\bibitem[{{Krolik} \& {Kriss}(2001)}]{krolik01}
{Krolik} J.H., {Kriss} G.A., Nov. 2001, ApJ, 561, 684

\bibitem[{{Krolik} et~al.(1981){Krolik}, {McKee}, \& {Tarter}}]{krolik81}
{Krolik} J.H., {McKee} C.F., {Tarter} C.B., Oct. 1981, ApJ, 249, 422

\bibitem[{{Lampton} et~al.(1976){Lampton}, {Margon}, \& {Bowyer}}]{lampton76}
{Lampton} M., {Margon} B., {Bowyer} S., Aug. 1976, ApJ, 208, 177

\bibitem[{{Mewe} et~al.(2003){Mewe}, {Porquet}, {Raassen} et~al.}]{mewe03}
{Mewe} R., {Porquet} D., {Raassen} A.J.J., et~al., Oct. 2003, In: {Brown} A.,
  {Harper} G.M., {Ayres} T.R. (eds.) The Future of Cool-Star Astrophysics: 12th
  Cambridge Workshop on Cool Stars, Stellar Systems, and the Sun, 1123--1128

\bibitem[{{Morini} et~al.(1987){Morini}, {Lipani}, \& {Molteni}}]{morini87}
{Morini} M., {Lipani} N.A., {Molteni} D., Jun. 1987, ApJ, 317, 145

\bibitem[{{Murphy} et~al.(1996){Murphy}, {Lockman}, {Laor}, \&
  {Elvis}}]{murphy96}
{Murphy} E.M., {Lockman} F.J., {Laor} A., {Elvis} M., Aug. 1996, ApJS, 105, 369

\bibitem[{{Ogle} et~al.(2004){Ogle}, {Mason}, {Page} et~al.}]{ogle04}
{Ogle} P.M., {Mason} K.O., {Page} M.J., et~al., May 2004, ApJ, 606, 151

\bibitem[{Page et~al.(2003)Page, Davis, \& Salvi}]{page03}
Page M.J., Davis S.W., Salvi N.J., 2003, MNRAS, 343, 1241

\bibitem[{{Page} et~al.(2003a){Page}, {Soria}, {Wu} et~al.}]{page03a}
{Page} M.J., {Soria} R., {Wu} K., et~al., Oct. 2003a, MNRAS, 345, 639

\bibitem[{{Perola} et~al.(2000){Perola}, {Matt}, {Fiore} et~al.}]{perola00}
{Perola} G.C., {Matt} G., {Fiore} F., et~al., Jun. 2000, A\&A, 358, 117

\bibitem[{Pounds et~al.(2001)Pounds, Reeves, O'Brien et~al.}]{pounds01}
Pounds K., Reeves J., O'Brien P., et~al., 2001, ApJ, 559, 181

\bibitem[{{Reynolds}(1997)}]{reynolds97}
{Reynolds} C.S., Apr. 1997, MNRAS, 286, 513

\bibitem[{{Singh} et~al.(1990){Singh}, {Westergaard}, {Schnopper}, {Awaki}, \&
  {Tawara}}]{singh90}
{Singh} K.P., {Westergaard} N.J., {Schnopper} H.W., {Awaki} H., {Tawara} Y.,
  Nov. 1990, ApJ, 363, 131

\bibitem[{Steenbrugge et~al.(2005)Steenbrugge, Kaastra, Crenshaw
  et~al.}]{steenbrugge05}
Steenbrugge K.C., Kaastra J.S., Crenshaw D.M., et~al., 2005, A\&A, 434, 569

\bibitem[{Stuhlinger et~al.(2005)Stuhlinger, Altieri, Esquej
  et~al.}]{Stuhlinger05}
Stuhlinger M., Altieri B., Esquej M.P., et~al., 2005, ESA SP-604, 2, 937

\bibitem[{{Tarter} et~al.(1969){Tarter}, {Tucker}, \& {Salpeter}}]{tarter69}
{Tarter} C.B., {Tucker} W.H., {Salpeter} E.E., Jun. 1969, ApJ, 156, 943

\bibitem[{{Weaver} et~al.(2001){Weaver}, {Gelbord}, \& {Yaqoob}}]{weaver01}
{Weaver} K.A., {Gelbord} J., {Yaqoob} T., Mar. 2001, ApJ, 550, 261

\bibitem[{Yaqoob et~al.(2003)Yaqoob, McKernan, Kraemer et~al.}]{yaqoob03}
Yaqoob T., McKernan B., Kraemer S.B., et~al., 2003, ApJ, 582, 105

\end{thebibliography}
\end{document}